\documentclass[twocolumn]{aastex62}
\newcommand{\be}{\begin{equation}}
\newcommand{\ee}{\end{equation}}
\def\lta{\,\raise 0.3 ex\hbox{$ < $}\kern -0.75 em
 \lower 0.7 ex\hbox{$\sim$}\,}
\def\gta{\,\raise 0.3 ex\hbox{$ > $}\kern -0.75 em
 \lower 0.7 ex\hbox{$\sim$}\,}

\DeclareMathAlphabet{\pazocal}{OMS}{zplm}{m}{n}

\usepackage{mathtools}
\DeclarePairedDelimiter\abs{\lvert}{\rvert}

\usepackage{amsmath}

\usepackage{xcolor, fontawesome, microtype}
\definecolor{twitterblue}{RGB}{64,153,255}
\definecolor{linkcolor}{rgb}{0.1216,0.4667,0.7059}

\usepackage{ mathrsfs }
\usepackage[T1]{fontenc}

\received{\today}
\revised{\today}
\accepted{\today}
\submitjournal{AJ}

\shorttitle{How the solar system is going to end}

\shortauthors{Zink et al. (2020)}
\begin{document}

\title{The Great Inequality and the Dynamical Disintegration of the Outer Solar System}

\correspondingauthor{Jon Zink}
\email{jzink@astro.ucla.edu}

\author[0000-0003-1848-2063]{Jon K. Zink}
\affiliation{Department of Physics and Astronomy, University of California, Los Angeles, CA 90095, USA}

\author{Konstantin Batygin}
\affiliation{Division of Geological and Planetary Sciences California Institute of Technology, Pasadena, CA 91125, USA}

\author{Fred C. Adams}
\affiliation{Physics Department, University of Michigan, Ann Arbor, MI 48109, USA}
\affiliation{Astronomy Department, University of Michigan, Ann Arbor, MI 48109} 

\begin{abstract}
Using an ensemble of N-body simulations, this paper considers the fate of the outer gas giants (Jupiter, Saturn, Uranus, and Neptune) after the Sun leaves the main sequence and completes its stellar evolution. Due to solar mass-loss -- which is expected to remove roughly half of the star's mass -- the orbits of the giant planets expand. This adiabatic process maintains the orbital period ratios, but the mutual interactions between planets and the width of mean-motion resonances (MMR) increase, leading to the capture of Jupiter and Saturn into a stable 5:2 resonant configuration. The expanded orbits, coupled with the large-amplitude librations of the critical MMR angle, make the system more susceptible to perturbations from stellar flyby interactions. Accordingly, within about 30 Gyr, stellar encounters perturb the planets onto the chaotic sub-domain of the 5:2 resonance, triggering a large-scale instability, which culminates in the ejections of all but one planet over the subsequent $\sim10$ Gyr. After an additional $\sim50$ Gyr, a close stellar encounter (with a perihelion distance less than $\sim200$ AU) liberates the final planet. Through this sequence of events, the characteristic timescale over which the solar system will be completely dissolved is roughly 100 Gyr. Our analysis thus indicates that the expected dynamical lifetime of the solar system is much longer than the current age of the universe, but is significantly shorter than previous estimates. 

\end{abstract}


\section{Introduction}

Understanding the long-term dynamical stability of the solar system constitutes one of the oldest pursuits of astrophysics, tracing back to Newton himself, who speculated that mutual interactions between planets would eventually drive the system unstable \citep{las96,las12}. \citet{lap99} and \citet{lag76} successfully challenged this perception by approximating the mutual interactions as perturbations, showing that, to leading order, the long-term evolution of all known solar system planets could be described via cyclic secular variations, thus analytically demonstrating the indefinite stability of the solar system. However, subsequent analyses by \citet{gau09} and \citet{lev56} showed that these approximations break down over sufficiently long time intervals, so that more complicated solutions are required. \citet{poi92} formalized this insight by proving that the full ``three-body problem'' could not be solved in closed form --- a barrier that has only recently been overcome with the advent of modern computing and N-body integration methods.

Unfortunately, even the most precise N-body simulations are only able to produce time-limited prognosis for the evolution of the solar system. Due to the chaotic nature of the planetary orbits, deterministic forecasting is impossible over sufficiently long timescales. In particular, the Lyapunov time for the inner terrestrial planets (Mercury, Venus, Earth, and Mars) is of order $\sim5$ Myr \citep{las89,sus92}, while the outer Jovian planets (Jupiter, Saturn, Uranus, and Neptune) appear chaotic with a Lyapunov time of order 10 Myr\footnote{The exact value for the Lyapunov exponent is strongly dependent on the initial conditions of the system and has be found to range from 5-20 Myr, using orbital parameters within the observational uncertainty.} \citep{las89,mur99,guz05,guz06}. Predictions on timescales significantly longer than these benchmark values are only meaningful in a statistical sense, recasting the question of the solar system's long-term fate as a probabilistic one.

Over the course of the last three decades, the question of whether or not the orbits of the solar system's eight planets can remain immutable has come into sharp focus, with state of the art simulations demonstrating that Mercury has a $\sim1\%$ chance of becoming unstable within the remaining main sequence (MS) lifetime of the Sun \citep{las94,las08,bat08,las09,zee15}. The mechanism for the onset of Mercury's instability is well understood: by virtue of locking into a linear secular resonance with the $g_5$ mode of the solar system's secular solution, Mercury's eccentricity can attain near-unity values, resulting in a collision with the Sun, or even Venus. Intriguingly, General Relativistic effects factor into this estimate, with ancillary apsidal precession providing a stabilizing influence on Mercury's orbit \citep{lit11,bou12,bat15b}. Within the context of this narrative, however, the remaining planets appear unaffected and are currently expected to remain stable for a lower limit of $10^{18}$ years, when diffusion arising from the overlapping mean motion resonance of Jupiter and Saturn are expected to decouple Uranus \citep{mur99}.

Although the estimate of \citet{mur99} addresses the intrinsic stability of the solar system, on sufficiently long timescales, extrinsic effects come into play. For example, stellar evolution, which is generally not considered in orbital stability studies, represents an important additional aspect of the problem. In particular, the Sun will undergo significant mass-loss over the next 7 Gyr, reducing the mass by roughly half, down to $0.54M_{\sun}$ \citep{sac93}. Over this time span, it is probable that Mercury, Venus, and Earth will be engulfed by the Sun, as its radius expands during the red-giant branch (RGB) phase of evolution \citep{ryb01,sch08}. This epoch thus marks the end of the three innermost planets. Although Mars, Jupiter, Saturn, Neptune, and Uranus will survive this phase of stellar evolution \citep{ver12}, they will experience a $\sim1.85$ increase in their semi-major axes \citep{jea24,ver11,ada13,ver16}. With their newly expanded orbits, the remaining planets are expected to remain stable for a minimum of 10 Gyr \citep{dun98}. However, the details of orbital evolution that unfolds on much longer timescales are less well characterized.

A distinct form of external forcing upon the solar system stems from stellar encounters. As the solar system traces through its Galactic orbit it will experience perturbations from passing stars, which will act to excite the orbits. If the solar system maintained its current orbital configuration, passing stars would liberate the planets over timescales of order $10^{14}$ yr \citep{dyson1979,ada1997}. However, the interaction cross-section for these gravitational encounters scale with the semi-major axis of the planet \citep{li15}. As a result, by accounting for stellar mass-loss and the inflation of the outer planet orbits, these encounters will become more influential. Given enough time, some of these flybys will come close enough to disassociate --- or destabilize --- the remaining planets. This paper examines the timescales and mechanisms that bring about the demise of the solar system. 
 
This paper is organized as follows. In Section \ref{sec:Methods}, we discuss the input parameters of our numerical study, including methods for simulating the evolution of the Sun itself and dynamical perturbations from stellar flybys. We then provide the results of our study and discuss the mechanisms for planetary disassociation in Section \ref{sec:result}. The paper concludes in Section \ref{sec:disc} with a summary of our results, a discussion of the significance of these findings, and the implications for free floating planets.

\section{Numerical Methods and System Parameters} 
\label{sec:Methods}

\subsection{The N-Body Simulation}
\label{sec:nBody}

The timescales of interest for this study greatly exceed the current age of the solar system. To carry out this simulation, we used the IAS15 high-order integrator \citep{rei15} as implemented in the {\tt Rebound} \citep{rei12} software package. This 15th order integrator uses a Gau{\ss}-Radau algorithm to numerically solve the equations of motion. Despite the high-order of this calculation, any use of finite time-steps introduces error, which can be characterized by deviations from the initial system energy ($\Delta E/E$). In an idealized system, IAS15 is capable of achieving $10^{-28}$ precision, but testing of the outer solar system produces a more realistic $\Delta E/E$ of $\sim10^{-19}$. However, this metric assumes that the system energy is conserved throughout the simulation and the current study considers the effects of extrinsic factors that stochastically modify the energy of the solar system itself.\footnote{We find a slow increase in energy $(\Delta E/E)\sim10^{-18}$ after a typical flyby interaction. This value is consistent with the finding of \citet{li15}, who did a similar flyby injection test. It is important to note that rare close encounter flybys have the ability to produce larger changes in the system energy.} Over sufficient time, the accumulated energy contributions from stellar flybys will dominate the $\Delta E/E$ metric. Without a meaningful measure of the systematic integration error, we rely on adaptive time-steps (see \citealt{qui97}), which automatically reduces the stepsize when the expected error exceeds machine precision ($\sim 10^{-16}$). Doing so, we acknowledge our inability to provide an accurate measure of this error, but expect $\Delta E/E$ to be of the order $\sim 10^{-16}$, the adaptive time-step limit.

\subsection{The Aging Solar System}

Since this study is focused on the long term effects of the outer solar system, we only include Jupiter, Saturn, Uranus and Neptune in our simulations. Although Mars is likely to survive the red giant phase of the Sun's evolution \citep{ver16}, its dynamical contribution is negligible and can be ignored in order to reduce computational costs. By focusing on the outer giant planets, we are only limited by the orbital resolution of Jupiter. Furthermore, the orbital period of Jupiter changes as the Sun loses mass, increasing the orbital period by a factor of $\sim3.4$ over the lifetime of the Sun. In order to take full advantage of this increased period, we break our simulation into two parts. The first part (Phase I) includes all of the stellar evolution, starting from the present epoch when the Sun is on the main-sequence, continuing through the red giant and mass-loss phases, and ending as the star becomes a white dwarf. The second part (Phase II) includes all of the subsequent temporal evolution, when the Sun has a fixed stellar mass and remains as a quiescent white dwarf. 

The Phase I epoch extends from today to the epoch of final mass-loss experienced at the end of the Sun's planetary nebula phase. During this phase, $46\%$ of the current solar mass will be ejected via stellar winds. To ensure that we appropriately account for the orbital effects due to this mass-loss, we implement the MESA \citep{pax19} solar mass evolution model with zero rotational velocity into our simulation. At each time-step ($\Delta t=216.63$ days, corresponding to 1/20 the orbital period of Jupiter) we update the mass of the Sun to reflect the value suggested by this model. To ensure our mass-loss is smooth, we linearly interpolate the stellar mass between the output profiles provided by the MESA simulations. Throughout most of the Phase I interval, the Sun retains a majority of its current mass and the orbits of the giant planets remain static. In the final $\sim1$ Myr of this period, however, the Sun loses $0.41M_{\Sun}$, and the semi-major axes ($a$) of the giant planets expand accordingly. As long as solar mass-loss occurs over many planetary orbits --- which is expected to be the case --- the orbits will experience adiabatic expansion (for a more in depth discussion of the adiabatic limit, see \citealt{ver11}).

The orbital expansion associated with solar mass-loss can be understood qualitatively as follows. A well-known result of classical perturbation theory (see e.g., \citealt{lic83}) is that an oscillator, subjected to slow parametric changes, will preserve the ratio of its energy ($E$) to its frequency ($n$). By analogy, the adiabatic invariant associated with a Keplerian orbit ($J$) has the form:
\be
J=\abs*{\frac{E}{n}}\approx\frac{GM_\star m/2a}{\sqrt{GM_\star/a^3}}\propto\sqrt{GM_\star a}\approx{\sl constant},
\ee
which also corresponds to the first Poincar{\'e} action (see \citealt{mor02}). Therefore, reduction in stellar mass ($M_\star$) will inflate the semi-major axis according to $a\sim1/M_\star$. To ensure this orbital expansion remains within the adiabatic limit, we have repeated this simulation, artificially slowing down the mass-loss rate (and hence orbital expansion) by a factor of 100, and find that the orbital parameters remain consistent with results obtained from the expected real-time expansion. 

After the epoch of solar mass-loss concludes, and the Sun remains as a $0.54M_{\Sun}$ white dwarf, we enter the second phase of our simulations. We now adjust the time-step to become 1/20th of the increased orbital period of Jupiter ($P_{orb}$ = 40.6 yr, $\Delta t=740.56$ days) at its new expanded orbit ($a=9.62$ AU) and lower stellar mass. This time-step increase allows for consistent orbital resolution and speeds up the simulations by a factor of 3.4.  

\subsection{Stellar Flyby Encounters}
\label{sec:flyby}
\begin{figure*}
\centering \includegraphics[height=7cm]{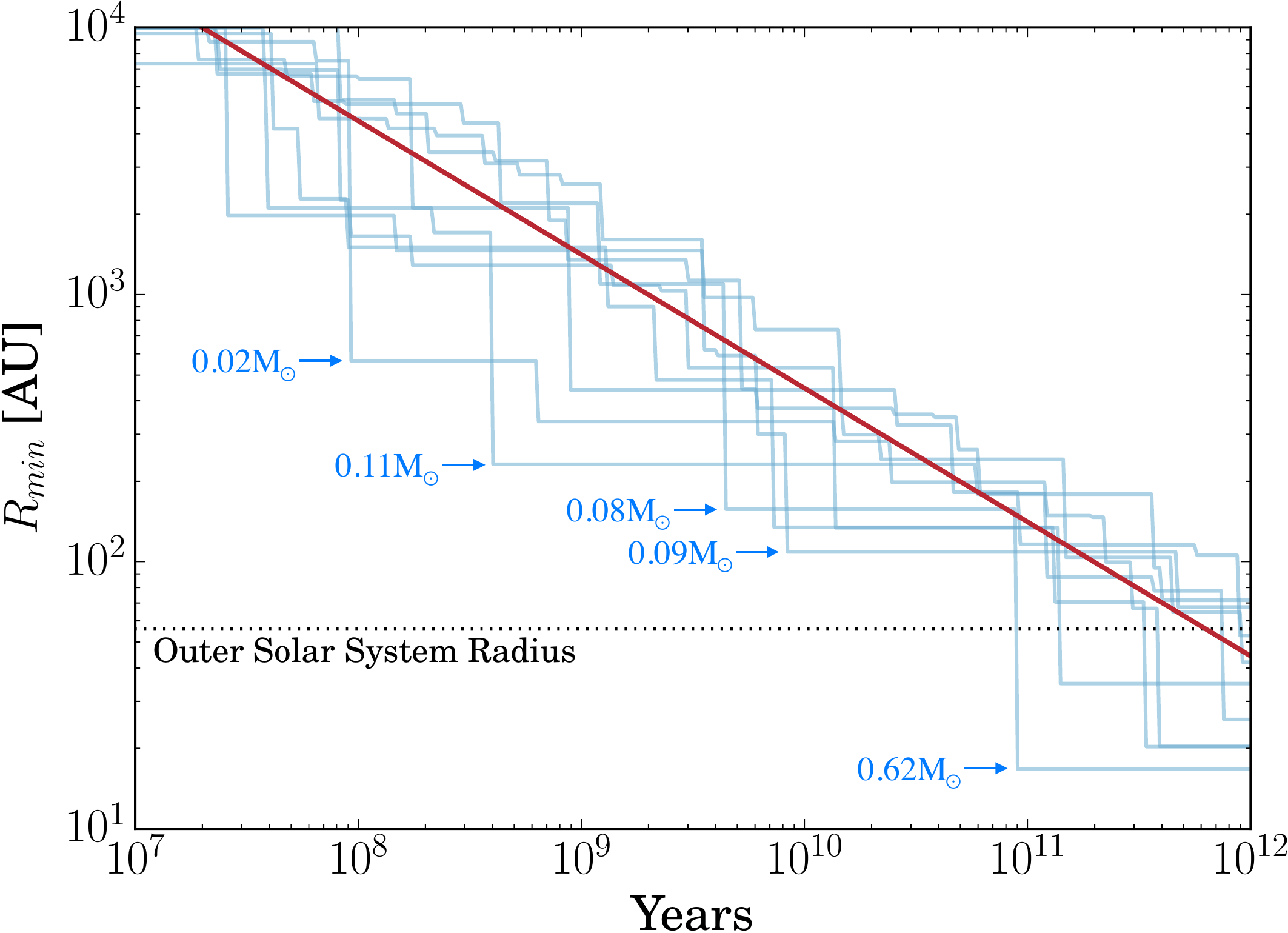}
\caption{The closest approach of flybys as a function of time. The blue lines show the minimum distances as a function of time realized in the 10 simulations and the red line indicates the expected value from Equation (\ref{eq:timeScale}). The outer solar system radius ($60$ AU) is represented by the semi-major axis of Neptune at the start of Phase II, when the Sun has reached its final (reduced) mass. The blue labels indicate the mass of the corresponding stellar encounter, highlighting that close encounters typically occur with low-mass stars. 
\label{fig:rmin}}
\end{figure*}

Throughout the simulations, we introduce stellar flybys which perturb the system, at a rate consistent with the current galactic environment of the solar system. To simulate these flybys, we draw a 10,000 AU sphere around the solar system and introduce incoming stars as described below. Only flybys within this sphere are resolved in our calculation. The expected rate ($\Gamma$) for the solar system to encounter passing stars can be written in the form: 
\be
\Gamma=\langle n_{\star}\rangle \pi B^2 \langle v\rangle \,,
\label{eq:gamma}
\ee
where $\langle n_{\star}\rangle$ is the local stellar number density (0.14 pc$^{-3}$; \citealt{mck15,bov12}), $B$ is the boundary of the interaction region (10,000 AU), and $\langle v\rangle$ is the expected local stellar velocity dispersion ($\sim40$ km/s; \citealt{bin08}). With these parameter values, the expected encounter rate $\Gamma$ is about $4.2\times10^{-8}$ stars/year. In other words, we expect a star to enter our sphere every 23 Myr. It is important to note that we use a static local galactic environment, which may change over the periods considered in the present study. Further discussion of this issue is provided in Section \ref{sec:galaxy}.

To simulate these flyby encounters we follow a procedure similar to that of \citet{hei86}. We first randomly select stellar masses from the initial mass function (using the form advocated by \citealt{kro01}) within a mass range of 0.08 -- $1M_{\Sun}$\footnote{We select an upper bound of $1M_{\Sun}$ as a conservative estimate of the effects of possible stellar encounters. In practice, we found invoking this bound had little effect on the overall outcome of our simulations. Our results thus provide an upper limit, in that the inclusion of more massive stars would reduce the time needed for planetary disassociation.} and assign a relative velocity ($v_{\inf}$) drawn from a Maxwell-Boltzmann distribution with a scale parameter $\langle v\rangle$. At each time-step, a new star is selected and the expected probability of encounter is calculated ($\Gamma\Delta t$; exchanging $v_{\inf}$ for $\langle v\rangle$) and an independently drawn random value ($\pazocal{R}[0,1]$) indicates whether the star will enter the sphere or not. If the star is permitted to dynamically engage with the solar system, the inclination for the flyby is drawn from a distribution of $\arcsin(\pazocal{R}[0,1])$. The angular orbital elements $\Omega$ and $\omega$ are uniformly drawn from a distribution of  $\pazocal{R}[-\pi/2,\pi/2]$. The impact parameter ($b$) of the flyby dictates the distance of closest approach ($r_p$) and is drawn from a distribution of $10000\sqrt{\pazocal{R}[0,1]}$ AU, as needed for a uniform sampling of the cross sectional area. For a hyperbolic orbit, the perihelion $r_p$ is related to the impact parameter according to $r_p$ = $\sqrt{a^2+b^2}-|a|$, where $a$ is the semi-major axis of the encounter. In other words, the relationship between $r_p$ and impact parameter $b$ takes the form 
\be
b^2= r_p^2 \left(1+\frac{2G(M+\langle M' \rangle)}{r_p\langle v\rangle^2}\right)\approx r_p^2 \,,
\label{perihelion} 
\ee
where $G$ is the gravitational constant, $M$ is the solar mass, and $\langle M' \rangle$ is the expected mass of the flyby star. The gravitational focusing term represents a small correction (only $\sim1\%$ for $r_p$ = 100 AU) which is negligible for nearly all the encounters considered in this study and can be ignored.

To show that our method of randomly selecting stellar encounters produces the correct distribution, we compare our simulations to the expected time $\tau$ at which a given minimum distance of closest approach ($R_{min}$) is achieved. The quantity $R_{min}$ is thus the minimum value of the perihelion $r_p$ experience by the solar system as a function of time. Using the simplification of equation (\ref{perihelion}), along with the relation $\tau=1/\Gamma$, we can calculate $R_{min}$ as a function of $\tau$: 
\be
R_{min}(\tau) = \left[ 
\langle n_{\star}\rangle \pi 
\langle v\rangle \tau \right]^{-1/2}\,.
\label{eq:timeScale}
\ee
Figure \ref{fig:rmin} presents the results of this comparison and shows that our sampling procedure replicates the expected time required to reach a given minimum distance of closest approach $R_{min}$. 

Although we insert stellar flybys throughout the full integration, we find that they have little effect on the solar system during Phase I. The tightly packed planets are effectively immune to these distant perturbations during the Sun's (relatively short) remaining lifetime as a main-sequence star. For example, the cross section for changing the orbit of Neptune enough to create significant disruption (specifically, so that it crosses the orbit of Uranus) is of order 1000 AU$^2$ \citep{laugh2000}, which corresponds to a distance of closest approach $R_{min}\sim20$ AU. Figure \ref{fig:rmin} shows that this value of $R_{min}$ is not achieved until a time $t\ge10^{12}$ yr, well beyond the time span of Phase I. Although more distant encounters are expected over the remainder of the Sun's main-sequence life, these perturbations will produce only small modifications to Neptune's eccentricity ($\Delta e\sim0.01$). Moreover, these minor perturbations will not have a significant impact on the inner gas giants. As result, the remainder of the paper focuses on the second part (Phase II) of our simulations.

\section{Results}
\label{sec:result}

We have carried out 10 simulations of the outer solar system's long-term evolution, with each run spanning $10^{12}$ years. While this ensemble of simulations does not constitute a large statistical sample, we find similar results in each case, indicating that the dynamical picture attained here is representative. In this section we discuss the findings of this study. In addition to determining the timescales for planetary ejection, we also elucidate the mechanisms that cause the solar system to dissolve.  

\subsection{Ejected planets}

\begin{figure*}
\centering \includegraphics[height=13cm]{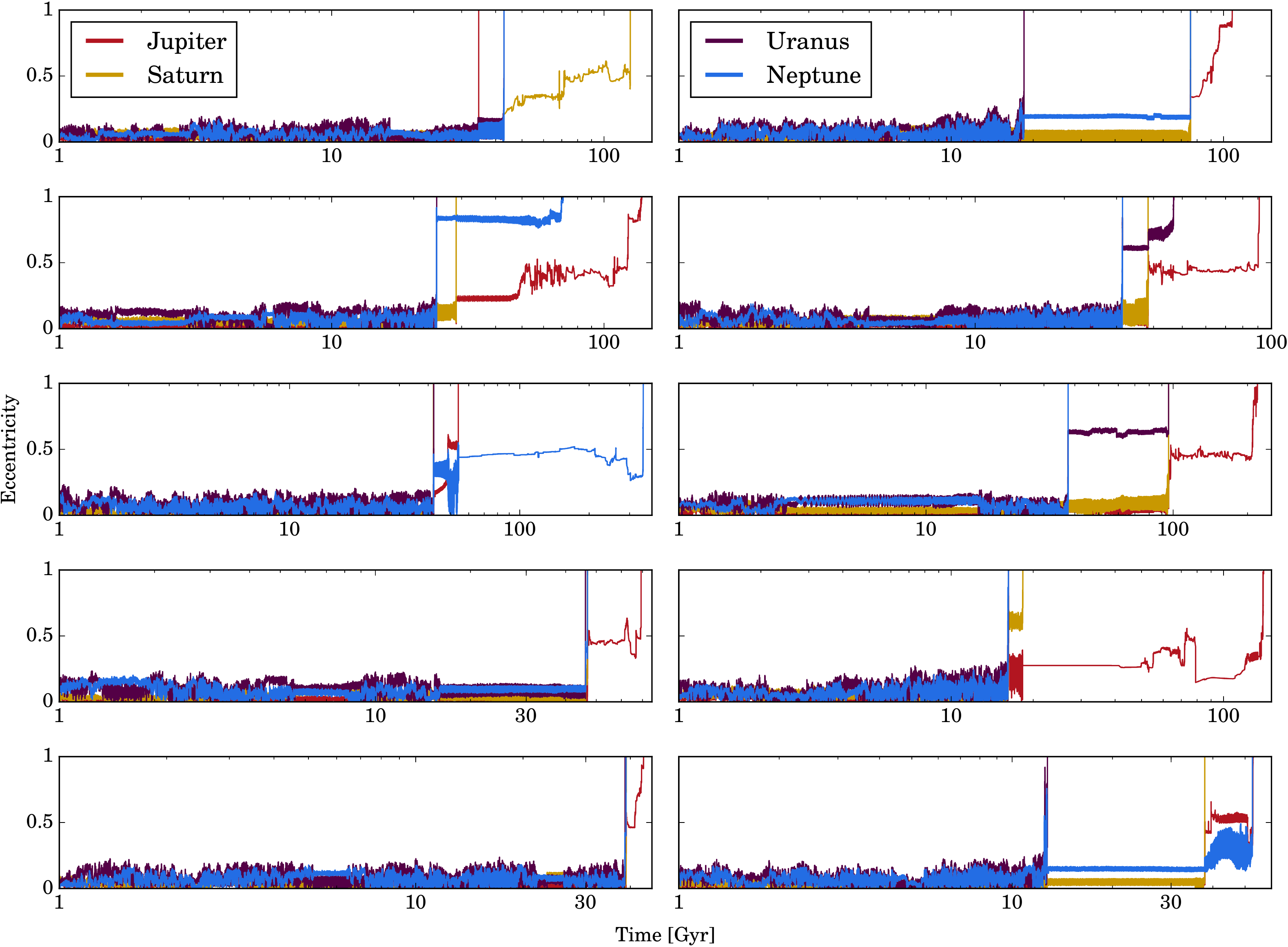}
\caption{Evolution of the outer solar system after solar mass-loss. Eccentricity of the gas giants is shown as a function of time. Given sufficient time, perturbations from a stellar encounter will drive the outer solar system planets unstable, ejecting all but one planet. An additional close stellar encounter is needed to remove the final planet. The time at which each planet is dissociated from the system is plotted in Figure \ref{fig:eject}.
\label{fig:allSim}}
\end{figure*}

After the Sun has completed its stellar evolution, including mass loss, the solar system will remain stable with the remaining planets orbiting with semi-major axes 1.85 times larger than their current values. As shown in Figure \ref{fig:allSim}, the eccentricities of all the planets remain low ($e\leq0.2$) during the first 10 Gyr of the Phase II era. This finding is consistent with the results of \citet{dun98}, who found no significant eccentricity growth during the initial onset of the Sun's white dwarf phase. However, our results begin to deviate beyond this 10 Gyr timescale.\footnote{\citet{dun98} acknowledged that calculations beyond 10 Gyr would require a more careful accounting of external stellar encounters, as carried out in this present study. } As depicted in Figure \ref{fig:rmin}, the occurrence of a stellar flyby with a perihelion less than 500 AU is likely within a 10 Gyr period, and such an encounter provides a significant perturbation to the system. Figure \ref{fig:allSim} indicates that these stellar encounters can  significantly increase the eccentricity of the planets, and even lead to the complete disassociation for many of the planets before the 45 Gyr benchmark for stability reported by \citet{dun98}.   

\begin{figure*}
\centering \includegraphics[height=7cm]{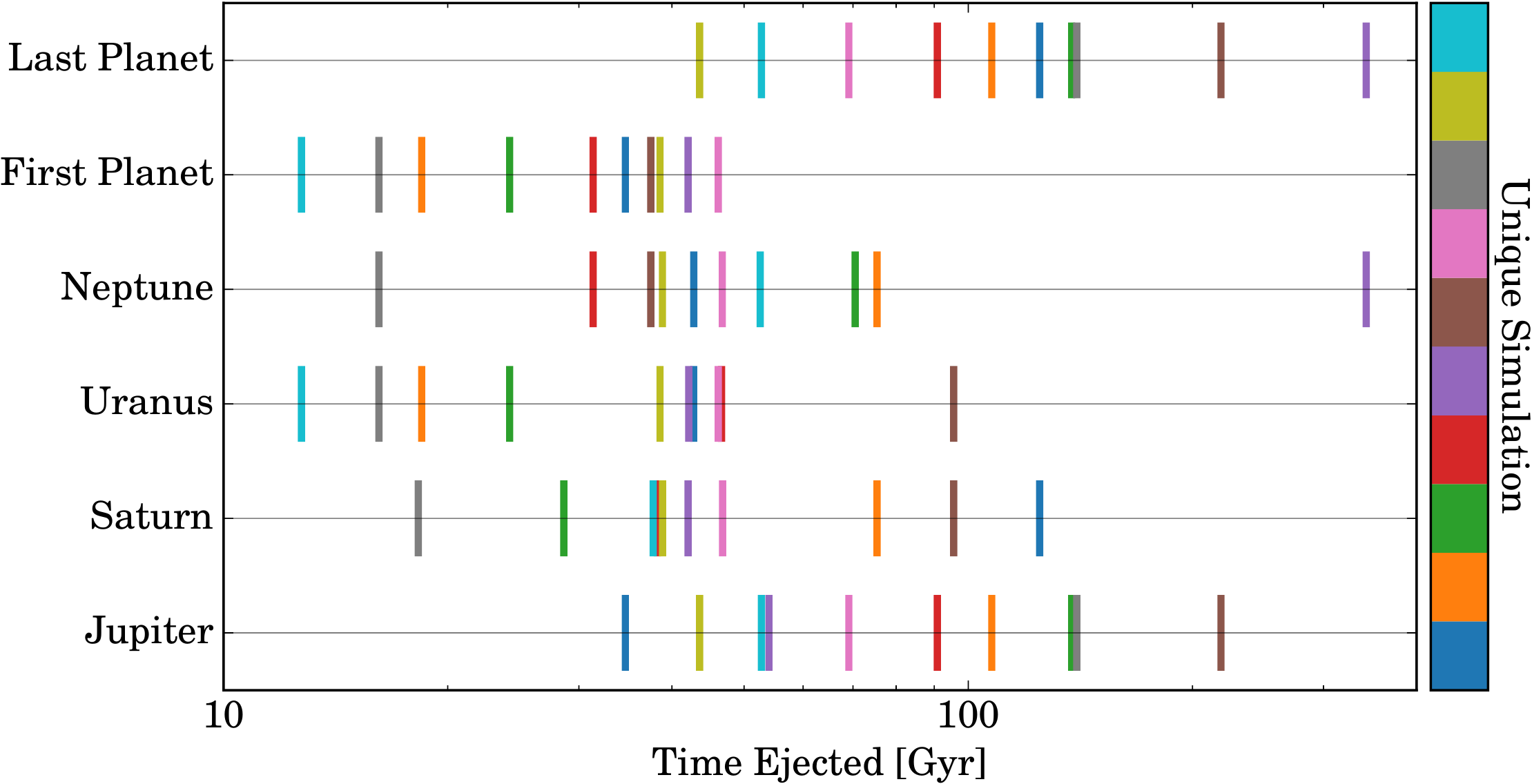}
\caption{The ejection time for each of the gas giants, after the Sun has become a white dwarf. We represent each of our 10 simulations with a different color. The First and Last Planet columns show the time at which the first and last planet was ejected from the system in a given simulation. In all cases the four gas giants were removed before the $10^{12}$ year limit of this study.
\label{fig:eject}}
\end{figure*}

In all 10 of our simulations, the four gas giants are ejected from the solar system within $10^{12}$ years, following the end of solar mass-loss. Figure \ref{fig:eject} presents the times at which each planet was removed from each of the simulations. The overall average ejection time is roughly 65 Gyr after mass-loss (72 Gyr from today). This timescale is far shorter than the $10^{18}$ yr lower bound predicted by \citet{mur99} for internal instability, and shorter than the timescale of $10^{14}$ yr predicted for external perturbations with a compact configuration \citep{dyson1979,ada1997}. Moreover, if we only consider the first planet ejected, we find an average ejection time of about 30 Gyr. In contrast, the last planet is ejected (on average) at a time of $\sim100$ Gyr. There is no definitive order in which the planets get removed, but typically the ice giants are removed first, with Uranus's ejection followed by Neptune, Saturn, and Jupiter, respectively. Usually the first three planets are all expelled within 5 Gyr of the first ejection. The remaining planet will then linger for an additional 50 Gyr before being removed from the system (see Section \ref{sec:last} for further discussion of this final planet). By accounting for the expanded Phase II planetary orbits and external perturbations from stellar flybys, we thus find a significantly reduced expected lifetime for planets to remain bound to the Sun. In the following sections, we discuss the mechanisms that drive this dissolution of the solar system. 

\subsection{Baseline Stability Considerations} 
\label{sec:criteria}

Arguably the simplest way to characterize the stability of a planetary system is to require sufficient separation in units of mutual Hill Radii \citep{cha96}. This condition is often written in the form 
\be
a_2 > a_1 + A R_H \,, 
\ee
where $a_k$ are the semi-major axes of adjacent planets and $A$ is a dimensionless constant that depends on the specific architecture of the system. Typically, one requires $A\gta10$ for stability of multiplanet systems \citep{pu15}, whereas smaller values are applicable for two-planet systems \citep{peti18}. The mutual Hill Radius $R_H$ is given by
\be
R_H = \left({a_1 + a_2 \over 2}\right) 
\left( {m_1 + m_2 \over 3M_\star}\right)^{1/3} \,,
\ee
where the $m_k$ denote the masses of adjacent planets and $M_\star$ is the mass of the Sun. As already discussed above, $M_\star$ varies over the history of the solar system, from $M_\star=1M_\odot$ at the present epoch down to $M_\star=0.54M_\odot$ after mass-loss. For a given pair of adjacent planets, we can thus define a dimensionless parameter $\Delta$ that provides a measure of system stability, i.e., 
\be
\Delta \equiv {2(a_2-a_1)\over a_2 + a_1 } 
\left({3M_\star \over m_2 + m_1} \right)^{1/3} \,. 
\ee
During the epoch of mass-loss, we expect the system to remain in the adiabatic regime so that $aM_\star\approx$ {\sl constant} as the Sun loses mass. As a result, the leading coefficient is invariant and the stability parameter scales according to $\Delta \sim M_\star^{1/3}$.  As the stellar mass decreases, the stability parameter also decreases, and the system becomes more unstable.

Applying the expected changes in stellar mass over the course of the Sun's lifetime, one finds the stability factor ($\Delta$) is reduced from today's value of 8 to 6 for the Jupiter/Saturn orbital spacing. Likewise, $\Delta$ is reduced from 14 to 11 for both the Saturn/Uranus and the Uranus/Neptune stability pairing.

\subsection{Jupiter and Saturn Resonance} 
\label{sec:JupSat}
Although the above discussion indicates that planet-planet interactions are expected to grow stronger due to solar mass-loss, the actual source of large-scale instability remains to be identified. Remarkably, our simulations suggest that Jupiter and Saturn's 5:2 near-commensurability may provide the mechanism that triggers this large-scale instability. 

\begin{figure*}
\centering \includegraphics[height=10cm]{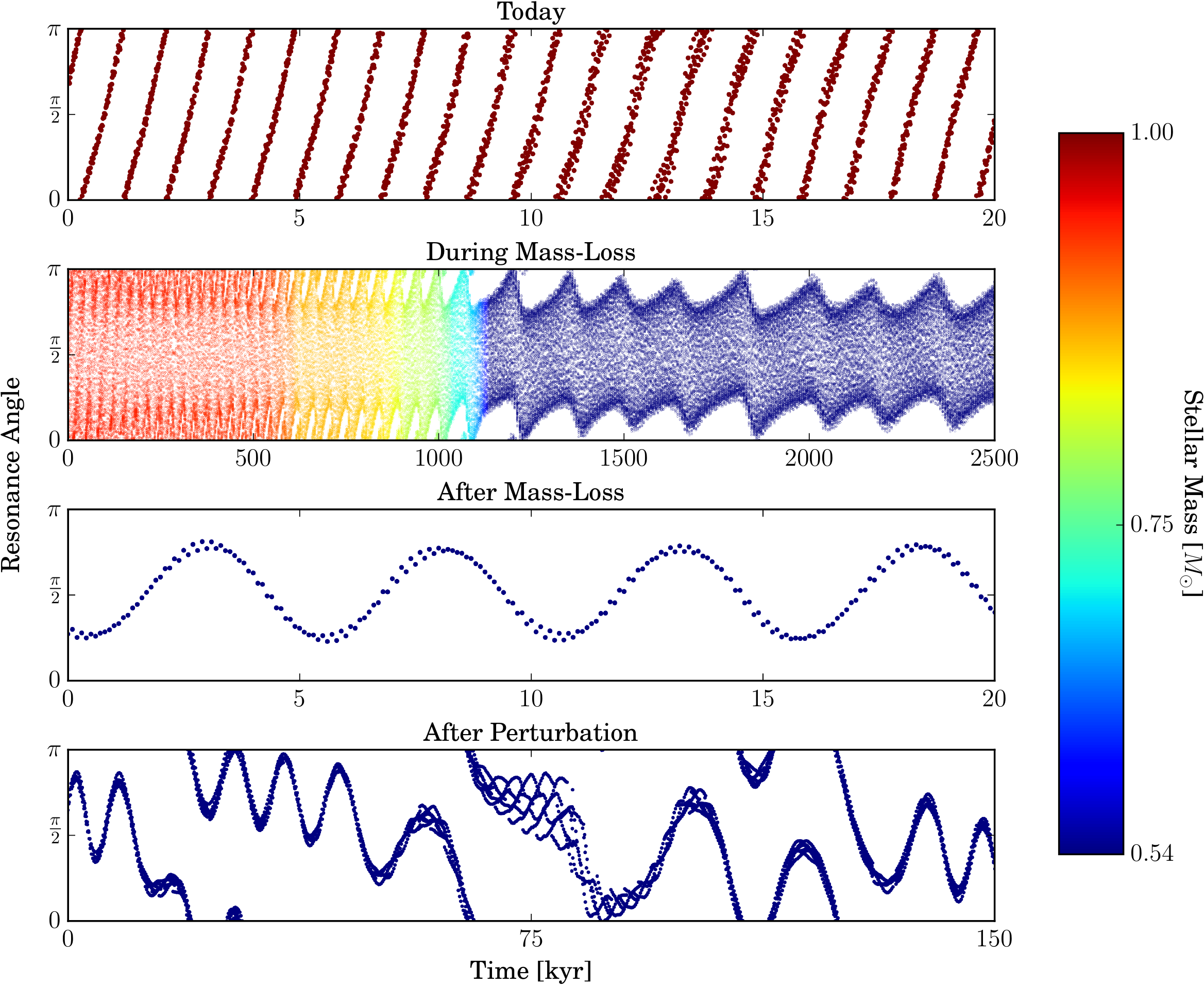}
\caption{The 5:2 mean-motion resonance (MMR) angle for Saturn and Jupiter ($5\lambda_{\textrm{Saturn}}-2\lambda_{\textrm{Jupiter}}-3\varpi_{\textrm{Saturn}})$ as a function of time. The data is colored based on the mass of the Sun at that given time-step. In the \textbf{Today} panel we show how the current angle circulates with a roughly $900$ year period. In the \textbf{During Mass-Loss} panel we show the adiabatic capture of Jupiter and Saturn into a 5:2 MMR. The \textbf{After Mass-Loss} panel shows the libration that occurs once the planets find their new expanded orbits after the solar mass-loss has been completed. In the \textbf{After Perturbation} panel we provide a sample of the chaotic circulation that transpires after being perturbed by a stellar flyby. This period of chaotic motion was followed by the ejection of Uranus 8 Myr later.   
\label{fig:resAngle}}
\end{figure*}

To make this argument, we consider the resonance angle $\phi = 5\lambda_{\textrm{Saturn}}-2\lambda_{\textrm{Jupiter}}-3\varpi_{\textrm{Saturn}}$, where $\lambda$ is the mean longitude and $\varpi$ is the longitude of pericentre for the respective orbits. Slow circulation of this angle indicates planets are near (but not in) mean-motion resonance (MMR). In other words, if the planets are not in MMR, the conjunction position will continuously move along the orbit. This behavior is characteristic of Jupiter and Saturn's present-day configuration. In the `Today' panel of Figure \ref{fig:resAngle}, we show the circulation of the Jupiter/Saturn resonant angle as seen today. The period of this circulation is about 900 years and is directly associated with the modulation in semi-major axes known as the ``Great Inequality'' (see Section \ref{sec:hist} for further discussion).

As the Sun loses mass, the semi-major axis ratio is conserved, but the width of the 5:2 MMR expands as $\sim M_\star^{-1/2}$ \citep{hen86}. This growth leads to the adiabatic capture of Jupiter and Saturn into the 5:2 MMR. In such a capture, the resonant angle will transition from circulation to libration. Our simulations indicate this transition occurs during the final $\sim1$ Myr of mass-loss (see the `During Mass-Loss' panel of Figure \ref{fig:resAngle}). Once the planets have been successfully captured, the resonant angle will execute bounded oscillations (see the `After Mass-Loss' panel of Figure \ref{fig:resAngle}). In isolation, this orbital configuration is stable on long timescales, as discussed in \cite{dun98}. However, the inflated orbits and weakened gravitational pull from the reduced solar mass render this new configuration more vulnerable to perturbations from stellar flybys.

The large amplitude ($\sim100\degr$) of the libration seen in the `After Mass-Loss' panel of Figure \ref{fig:resAngle} is indicative of a MMR system near the separatrix (i.e., the MMR boundary, where the resonant angle changes from libration to circulation). For roughly 30 Gyr after solar mass-loss, the system remains stable. Given sufficient time, however, a close encounter ($R_\textrm{min}<500$ AU) from a stellar flyby will create a perturbation large enough to perturb Jupiter and Saturn into the chaotic region of the 5:2 MMR (see \citealt{mor02}). This event triggers chaotic diffusion as shown in the `After Perturbation' panel of Figure \ref{fig:resAngle}, leading to large-scale instabilities in the outer solar system. In most cases the Jupiter/Saturn resonance angle will repeatedly switch from circulation to libration and back, pumping up the eccentricity of Uranus, Neptune, and Saturn until they are ejected from the solar system. This process takes place over $\sim10$ Gyr after the onset of accelerated chaos. Jupiter is usually the last planet remaining, but this ordering is not the only possible outcome. In one case Saturn was the final remaining planet and in another Neptune survived the tumultuous instability of the inner gas giants. However this large-scale instability played out, one planet remains orbiting the Sun for an extended period of time in most of the simulations.  
 
\subsection{The Last Planet Standing}
\label{sec:last} 

In all but one simulation, we found that a single planet remains orbiting the Sun for about 50 Gyr. (In the exceptional case, orbital diffusion left Jupiter with an eccentricity greater than $e\sim0.9$, leading to a swift ejection, 200 Myr later.) With an absence of additional planets, the surviving planet lacks a direct mechanism to attain positive energy. The only remaining source of energy exchange is through interactions with passing stars. Significantly, the large-scale instability that led to the ejection of the other three gas giants leaves the final planet with a heightened eccentricity (typically in the range $e\sim0.2-0.5$). As shown by \citet{li15}, the dynamical cross-section required for ejection is an exponential function of eccentricity. As a result,  the expected timescale for ejection of the post-instability gas giant is decreased by roughly a factor of two (compared to the planetary orbit before the onset of instability). 

Since flyby encounters are rare (entering the 10,000 AU sphere once every 23 Myr), and most interactions will have small dynamical effects on the remaining planet, the process of ejection can in principle occur steadily (e.g., through  incremental increases in orbital eccentricity and semi-major axes). On the other hand, given sufficient time, it is also possible that an extremely close encounter will independently liberate the final planet. The underlying mechanism for the removal of the final planet thus represents a competition between these two processes. In other words, will the final planet be ejected by a single major event or many small energy exchanges?

To intuitively understand the timescales over which these possible outcomes take place, we can crudely describe the process as a random walk. The follow derivation is merely an order of magnitude calculation, similar to the dynamical relaxation time derived by \cite{bin08}. The cumulative change in velocity of $\Delta U$ is given by 
\be
\Delta U \sim \Delta u \sqrt{N}\,,
\label{eq:rand}
\ee
where $\Delta u$ is the change in the velocity of the orbiting planet from a single interaction and $N$ represents the number of interactions. The planet must attain a positive energy in order to be disassociated from the system. To leading order, this requirement can be written in the form 
\be
\frac{(\Delta U)^2}{2}\approx \frac{GM_{\star}}{2a}.
\ee
Assuming all interactions follow the impulse approximation, we can express $\Delta u$ as:
\be
\Delta u = \frac{2GM_{\star}}{vb}\,,
\label{eq:impact}
\ee
where $v$ is the velocity of the flyby star and $b$ is the impact parameter for the stellar encounter. In reality, most interactions are much weaker and produce smaller changes in the planets orbital velocity. As a result, this calculation will provide a limiting constraint. Recalling Equation (\ref{eq:gamma}), we note that the number of interactions can be calculated as $N=\Gamma t$, where $t$ is the expected time needed to achieve $N$ interactions. Finally, we can use Equation (\ref{eq:rand}) to solve for the time required to eject the planet, i.e., 
\be
t=\frac{v^2}{4\pi G M_{\star}a \langle n_{\star}\rangle \langle v\rangle }\,.
\label{eq:tscale}
\ee
Note that this equation is independent of the flyby impact parameter.\footnote{A similar calculation can be achieved through a random walk process of the orbital energy, under the assumption of the impulse approximation, culminating in a solution that deviates by only a factor of 4 from Equation \ref{eq:tscale}.} Under the assumption that the impulse approximation is valid, we find that a single planet is equally likely to be ejected by a series of distant encounters or by a single close encounter flyby. However, this approximation breaks down for adiabatic interactions, where the the gravitational interaction timescale ($T_{\textrm{enc}}$) is much greater than the planet's orbital period ($P$),
\be
T_{\textrm{enc}}\sim \frac{2b}{v} \gg P.
\ee
In typical cases, the star will enter the 10,000 AU sphere of influence with a velocity of order 40 km/s and an impact parameter of order $7000$ AU. These values indicate an interaction time $T_{\textrm{enc}}\sim$ 1800 yr, which is nearly five times greater than Neptune's Phase II period ($350$ years). Most interactions that occur after the large-scale instability has isolated a single planet will thus be adiabatic. In these cases, where perturbations are effectively secular, semi-major axis growth will be suppressed \citep{bat20} relative to predictions from the impulse approximation. In other words, the distant encounters will be weaker than the limiting case of Equation (\ref{eq:impact}). As a result, a single extreme encounter is more likely to liberate the final planet than the accumulated effects of many distant perturbers. Nonetheless, both the cumulative effects of many distant encounters and the impact of rare close encounters are likely to play a role. 

\begin{figure*}
\centering \includegraphics[height=8cm]{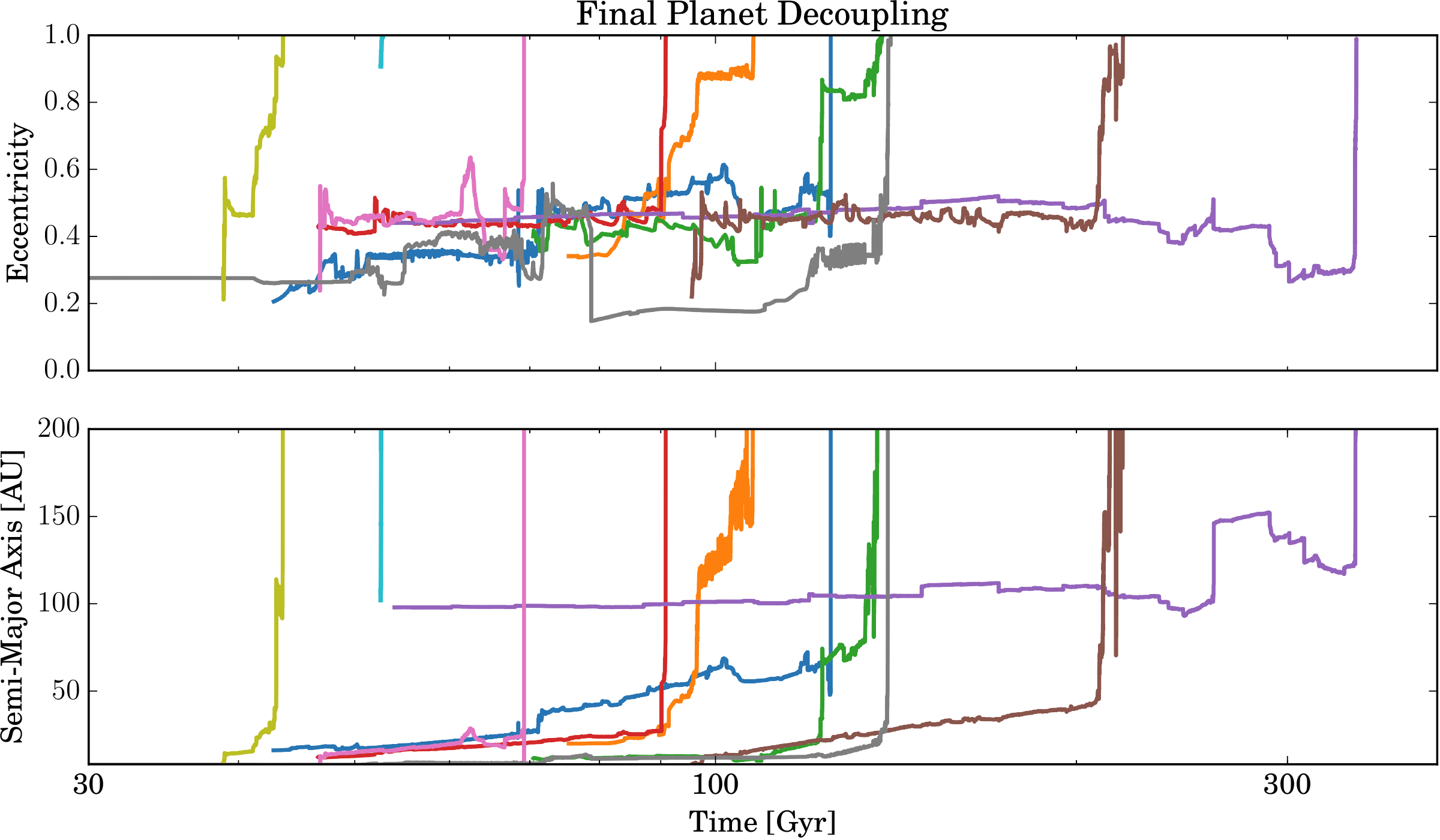}
\caption{The eccentricities and semi-major axes of the final planet, which remains bound after all of the other planets have been ejected from the system. The slow rise in eccentricity and semi-major axis are due to energy exchanges with multiple stellar flybys, but an important mechanism for disassociation of the final planet is an extremely close flyby encounter (with $R_\textrm{min}\le200$ AU).  
\label{fig:ecc}}
\end{figure*}

In Figure \ref{fig:ecc} we show how the eccentricity and semi-major axis of our simulated final planet changes before eventual disassociation. From the onset of isolation, nearly all cases show slow orbital diffusion due to weak gravitational interactions. Providing consistency with the above discussion, four of the simulations show the eventual disassociation of the final planet by a single extremely close flyby encounter  (with $R_\textrm{min}<200$ AU). In the remaining simulations, a close encounter significantly modifies the orbit, making the planet far more susceptible to perturbations from subsequent stellar flybys. After surviving this initial interaction, the final planet is ejected within a few Gyr. In both scenarios, we find that both major and minor energy exchanges play a role in the removal of the final planet. However, the majority of the liberating energy comes from a single close encounter.

\subsection{Caveats}

\subsubsection{Binary Encounters}

Within the current study we assume all flybys are single star encounters, in reality, however, observations indicate that roughly half of all nearby stars are part of a binary system (e.g., \citealt{Raghavan2010}; \citealt{Duchene2013}). Previous dynamical studies, considering these binary interactions, have found that the two stars will appear as independent perturbers when the relative flyby velocity ($v$) of the binary system is sufficiently large, increasing the interaction cross-section by roughly a factor of two \citep{laugh2000, li15}. However, when the orbital velocity of the interacting planet and the velocity between binary members are near the relative flyby velocity, the gravitational influence is enhanced. In this scenario, the extended interaction time allows motion from the binary system orbit to increase the gravitational cross-section (by a factor greater than 2), thereby increasing the likelihood of a significant perturbation.

The 40 km/s expected flyby velocity used for this study is quite large compared to the orbital speeds of the gas giants (Jupiter's orbital speed today is about 13 km/s and decreases to 9 km/s after solar mass-loss). Therefore, this effect would only be applicable for slow close encounter flybys, which alone would be completely disruptive to the orbiting planets. However, the shear number of binary systems in existence would increase the dynamical cross-section for the interactions. By choosing to exclude these binary encounters, we are making a conservative estimate for the lifetime of the future solar system. In other words, the effect of including binary flybys would further reduce this expected lifetime.\footnote{Note that if the binary fraction is 1/2, then the factor by which the cross section is increased would be 3/2.} 

\subsubsection{Galactic Evolution}
\label{sec:galaxy}

Over the timescales considered in this study, the solar system may undergo radial migration through the Galaxy, encountering regions of differing stellar density and velocity dispersion \citep{hal15}. However, the magnitude and direction of this migration remains an active area of discussion (e.g., \citealt{ros08} and \citealt{mar17}). As one example, the latter authors find that the encounter rate could vary 
by a factor of $\sim3$ if the solar system migrates outwards versus 
inwards. Continued star formation can also increase the stellar density. Acting in the opposite direction, the galactic disk tends to increase its velocity dispersion and hence its scale height over comparable timescales. In addition, the Milky Way is likely to collide with the Andromeda Galaxy over the next Hubble time, or two, again modifying the local galactic environment \citep{bin08}. These changes will impact the rate and velocity of stellar encounters, but accurately estimating these changes remains difficult and is beyond the scope of this present work. In this study, the current local stellar density and velocity dispersion are considered fixed throughout our simulations. One should keep in mind, however, that a more precise accounting of these future changes could modify our results.

\section{Conclusion}
\label{sec:disc}

Using a suite of long-term simulations of the solar system, that account for solar mass-loss and extrinsic forcing from passing stars, we have demonstrated that the expected dynamic lifetime of the outer planets is of order 100 Gyr, --- significantly shorter than previous estimates. Moreover, we have identified the specific dynamical pathway responsible for the onset of the solar system's final large-scale instability. The narrative emerging from our calculations can be summarized as follows. As solar mass-loss unfolds, the planetary orbits expand adiabatically and maintain their period ratios. At the same time, mutual planet-planet interactions grow stronger for two reasons: The separation of planetary orbits in units of mutual Hill radii decreases, and the width of mean-motion resonances (MMR) expand in concert. The process culminates in the adiabatic capture of Jupiter and Saturn into the 5:2 MMR, giving way to a period of stable resonant motion, one that is characterized by large-amplitude librations of the associated critical angle. In time, however, stellar flybys perturb the giant planets onto the chaotic sub-domain of the 5:2 MMR. Correspondingly, orbital diffusion ensues, leading to eventual orbit crossings, and ejection of the outer planets. This process is responsible for the ejection of all but a single remaining planet, which continues to orbit with an eccentricity that has been excited by the aforementioned large-scale instability. The planet's increased eccentricity and expanded semi-major axis enhances the subsequent dynamical interactions due to stellar encounters, which lead to the expeditious disassociation of the final gas giant. 

\subsection{Historical Context}
\label{sec:hist}

It is worth noting that speculation regarding large-scale instabilities within the solar system that are driven by the 5:2 resonant interaction between Jupiter and Saturn dates back to the work of \citet{new87,new13,new26}, as well as the pioneering development of perturbation theory by \citet{lap99}. Accordingly, let us contextualize our results against the back-drop of this remarkable saga.

The earliest data on the orbits of Jupiter and Saturn date back to Ptolemy in 228BC (Saturn) and 240BC (Jupiter) and were not recorded again until 1590 (nearly 1800 years later) when Tycho Brahe measured their orbital positions (see the account of \citealt{las96}). When Kepler set about mapping the elliptical orbits of our solar system, in 1625, he was unable to reconcile the data collected by Ptolemy, using the model derived from Tycho Brahe's observations. He noted that Ptolemy's observation required a slower mean motion for Jupiter and a faster mean motion for Saturn \citep{wil85}. In an effort to understand this discrepancy, Halley (1687) extrapolated the semi-major axes of planets over this period of time and determined Jupiter was slowly moving inward while Saturn's orbit was moving outward \citep{las96}. At face value, the data implied that given sufficient time Jupiter would collide with the Sun and Saturn would expand out into deep space.   

When \citet{new87} announced the universal law of gravity, he indicated that such deviations from the invariant elliptical orbits were due to mutual gravitational interactions. His belief was that such perturbations, lacking divine intervention, would eventually lead to the demise of the solar system \citep{las96}. However, Newton was not able to directly explain these observations since perturbation theory had not yet been developed.

\citet{lap99} finally explained this phenomenon, which became known as the ``Great Inequality'', by showing this was not a secular effect, but rather a periodic modulation of Jupiter and Saturn's semi-major axes by the 5:2 near resonance. As seen in the `Today' panel of Figure \ref{fig:resAngle}, this nearly $900$ year circulation leads to a cyclical expansion and contraction of the Jovian and Saturnian orbits, and appears to be stable for the remainder of the Sun's life as a main sequence star.

Our simulations show that the mechanism responsible for the eventual disintegration of the outer solar system is keenly related to the ``Great Inequality''. That is, the expanded orbits during Phase II (after solar mass-loss) allow the planets to be captured in a 5:2 MMR, producing large-scale instability when perturbed by stellar flybys. As a result, Jupiter and Saturn appear to be responsible for the ultimate demise of the solar system, only at a much later time than originally prophesied by Newton.    

\subsection{Free Floating Planets}

Once liberated, the outer solar system planets will independently roam through the Galaxy, becoming Free Floating Planets (FFP). Current estimates, using gravitational microlensing, suggest there are less than about 0.25 FFPs for every main-sequence star in the Galaxy \citep{prz17}. The exact origin of these abundant planets remains unclear, but it is probable that a large fraction of these FFPs were once bound to a host star and disassociated via dynamical instability \citep{for03}.

Here, we have shown that the outer gas giants of the solar system will eventually be ejected and contribute to this reservoir of FFPs. Given that all stars will experience some amount of mass-loss over their lifetimes, it is likely that other planetary systems --- that survive stellar evolution --- will also eventually experience large-scale instability (e.g. \citealt{ver11}). We can therefore conclude that as the Galaxy, and the stars that reside in it, continue to age the number of FFPs will increase as a function of time.

$\,$

\noindent
{\bf Acknowledgment:}
We would like to thank Kevin Hayakawa for his instructive discussion of long-term N-body simulations. The simulations described here were performed on the UCLA Hoffman2 shared computing cluster and used resources provided by the Bhaumik Institute. K.B. is grateful to the David and Lucile Packard Foundation and the Alfred P. Sloan Foundation for their generous support. The work of F.C.A. is supported by in part by NASA grant number NNX16AB47G and by the University of Michigan.

\bibliography{bib.bib}\setlength{\itemsep}{-2mm}

\end{document}